\documentclass[12pt,a4paper]{article}
\usepackage{a4p,graphicx,xspace}

\newcommand{\beq}{\begin{equation}}
\newcommand{\eeq}{\end{equation}}
\newcommand{\bea}{\begin{eqnarray}}
\newcommand{\eea}{\end{eqnarray}}
\newcommand{\bseq}{\begin{subequations}}
\newcommand{\eseq}{\end{subequations}}
\newcommand{\bsea}{\begin{subeqnarray}}
\newcommand{\esea}{\end{subeqnarray}}
\newcommand{\bit}{\begin{itemize}}
\newcommand{\eit}{\end{itemize}}
\newcommand{\ben}{\begin{enumerate}}
\newcommand{\een}{\end{enumerate}}
\newcommand{\bfig}{\begin{figure}}
\newcommand{\efig}{\end{figure}}
\newcommand{\btab}{\begin{table}}
\newcommand{\etab}{\end{table}}
\newcommand{\im}[1]{\ensuremath{#1}\xspace}       
\newcommand{\imx}[1]{\ensuremath{#1}\xspace}       
\newcommand{\etal}{{\em et al.}}

\newcommand{\eV}{\imx{\mathrm{e\kern -0.1em V}}}
\newcommand{\MeV}{\imx{\mathrm{Me\kern -0.1em V}}}
\newcommand{\GeV}{\imx{\mathrm{Ge\kern -0.1em V}}}
\newcommand{\TeV}{\imx{\mathrm{Te\kern -0.1em V}}}

\newcommand{\IM}{\im{\mathrm{I} \kern -0.15 em \mathrm{m}}}         
\newcommand{\RE}{\im{\mathrm{R} \kern -0.15 em \mathrm{e}}}         

\newcommand{\Pchi}{\im{{\raise5pt\hbox{$\chi$}}}}
\newcommand{\Pe}{\im{\mathrm{e}}}            
\newcommand{\Pm}{\im{\mu}}
\newcommand{\Pt}{\im{\tau}}
\newcommand{\Pn}{\im{\nu}}
\newcommand{\Pl}{\im{\ell}}
\newcommand{\Pp}{\im{\mathrm{p}}}

\newcommand{\Pc}{\im{\mathrm{c}}}
\newcommand{\Pb}{\im{\mathrm{b}}}
\newcommand{\PT}{\im{\mathrm{t}}}
\newcommand{\Pq}{\im{q}}
\newcommand{\Pf}{\im{f}}
\newcommand{\Phad}{\im{\mathrm{had}}}

\newcommand{\PW}{\im{\mathrm{W}}}             

\newcommand{\PZ}{\im{\mathrm{Z}}}
\newcommand{\PH}{\im{\mathrm{H}}}

\newcommand{\MW}{\im{M_{\PW}}}
\newcommand{\MZ}{\im{M_{\PZ}}}
\newcommand{\MH}{\im{M_{\PH}}}

\newcommand{\MT}{\im{M_{\PT}}}

\newcommand{\G}{\im{\Gamma}}                  
\newcommand{\GW}{\im{\G_{\PW}}}               

\newcommand{\A}{\im{\mathrm{A}}}

\newcommand{\Al}{\im{\A_{\Pl}}}

\newcommand{\Afb}{\im{\A_{\mathrm{fb}}}}

\newcommand{\Afbzb}{\im{\Afb^{0,\Pb}}}
\newcommand{\Afbzc}{\im{\Afb^{0,\Pc}}}

\newcommand{\swsqeffl}{\sin^2\theta_{\mathrm{eff}}^{\mathrm{lept}}}

\newcommand{\Pee}{\im{\Pe^+\Pe^-}}
\newcommand{\Pmm}{\im{\Pm^+\Pm^-}}
\newcommand{\Ptt}{\im{\Pt^+\Pt^-}}
\newcommand{\Pll}{\im{\Pl^+\Pl^-}}
\newcommand{\Ppp}{\im{\Pp\overline{\Pp}}}

\newcommand{\Pnn}{\im{\Pn\overline{\Pn}}}
\newcommand{\Pqq}{\im{\Pq\overline{\Pq}}}
\newcommand{\PTT}{\im{\PT\overline{\PT}}}
\newcommand{\Pbb}{\im{\Pb\overline{\Pb}}}

\newcommand{\Pff}{\im{\Pf\overline{\Pf}}}
\newcommand{\PWW}{\im{\PW^+\PW^-}}
\newcommand{\PZZ}{\im{\PZ\PZ}}

\newcommand{\Peeee}{\im{\Pee \kern -0.35em \rightarrow\Pee}}
\newcommand{\Peemm}{\im{\Pee \kern -0.35em \rightarrow\Pmm}}
\newcommand{\Peett}{\im{\Pee \kern -0.35em \rightarrow\Ptt}}
\newcommand{\Peell}{\im{\Pee \kern -0.35em \rightarrow\Pll}}
\newcommand{\Peenn}{\im{\Pee \kern -0.35em \rightarrow\Pnn}}
\newcommand{\Peeqq}{\im{\Pee \kern -0.35em \rightarrow\Pqq}}
\newcommand{\Peehad}{\im{\Pee \kern -0.35em \rightarrow\Phad}}
\newcommand{\Peeff}{\im{\Pee \kern -0.35em \rightarrow\Pff}}
\newcommand{\PeeTT}{\im{\Pee \kern -0.35em \rightarrow\PTT}}
\newcommand{\PeeWW}{\im{\Pee \kern -0.35em \rightarrow\PWW}}
\newcommand{\PeeZZ}{\im{\Pee \kern -0.35em \rightarrow\PZZ}}
\newcommand{\aqcd}{\im{\alpha_S}}

\newcommand{\GF}{\im{G_{\mathrm{F}}}}

\newcommand{\dalhad}{\im{\Delta\alpha^{(5)}_{had}}}

\begin{document}

\noindent                
{\Large
 $\phantom{0}$        \hfill UCD-PHYC/070901\\[1mm]
 $\phantom{0}$        \hfill arXiv:0709.3744 [hep-ex]\\[2mm]
 $\phantom{0}$\bf     \hfill 24 September 2007\\[1mm]
}
\begin{center}

\vskip 4cm

\vskip 1cm
{\Huge\bf Combined Electroweak Analysis \\}
\vskip 1cm
{\Large {\bf Martin W. Gr\"unewald}\\[20pt]
        UCD School of Physics\\
        University College Dublin\\
        Belfield, Dublin 4\\
        Ireland\\}
\vfill
{\bf Abstract} \\[10pt]
\end{center}
{
  
  Recent developments in the measurement of precision electroweak
measurements are summarised, notably new results on the mass of the
top quark and mass and width of the W boson.  Predictions of the
Standard Model are compared to the experimental results which are used
to constrain the input parameters of the Standard Model, in particular
the mass of the Higgs boson.  The agreement between measurements and
expectations from theory is discussed.

}
\vskip 1cm
\begin{center}
\em{ Invited talk presented at the EPS HEP 2007 conference,\\
  Manchester, England, July 19th to 25th, 2007}
\end{center}

\clearpage

\setcounter{page}{2}

\section{Introduction}

On the level of realistic observables such as measured cross sections,
ratios and asymmetries, the set of electroweak precision data consists
of over thousand measurements with partially correlated statistical
and systematic uncertainties. This large set of results is reduced to
a more manageable set of 17 precision results, so-called pseudo
observables, in a largely model-independent procedure, by the LEP and
Tevatron Electroweak Working Groups.  The pseudo observables updated
for this conference are summarised. Using in addition external
constraints on the hadronic vacuum polarisation at the Z pole and
``constants'' such as the Fermi constant $\GF$, analyses within the
framework of the Standard Model (SM) are performed~\cite{LEPEWWG}.

\section{Measurements}

More than 3/4 of all pseudo observables arise from measurements
performed in electron-positron collisions at the Z resonance, by the
SLD experiment and the LEP experiments ALEPH, DELPHI, L3 and OPAL.
The Z-pole observables are: 5 observables describing the Z lineshape
and leptonic forward-backward asymmetries, 2 observables describing
polarised leptonic asymmetries measured by SLD with polarised beams
and at LEP exploiting tau polarisation, 6 observables describing b-
and c-quark production at the Z pole, and finally the inclusive
hadronic charge asymmetry.  The Z-pole results and their combinations
are final and published since last year~\cite{Z-POLE}.  The remaining
pseudo observables are: the mass and total width of the W boson
measured by CDF and {D\O} at the Tevatron and by the four LEP-II
experiments, and the top-quark mass measured by the Tevatron
experiments only.

\begin{figure}[htbp]
\begin{center}
$ $\vskip -1cm
\includegraphics[width=0.49\linewidth]{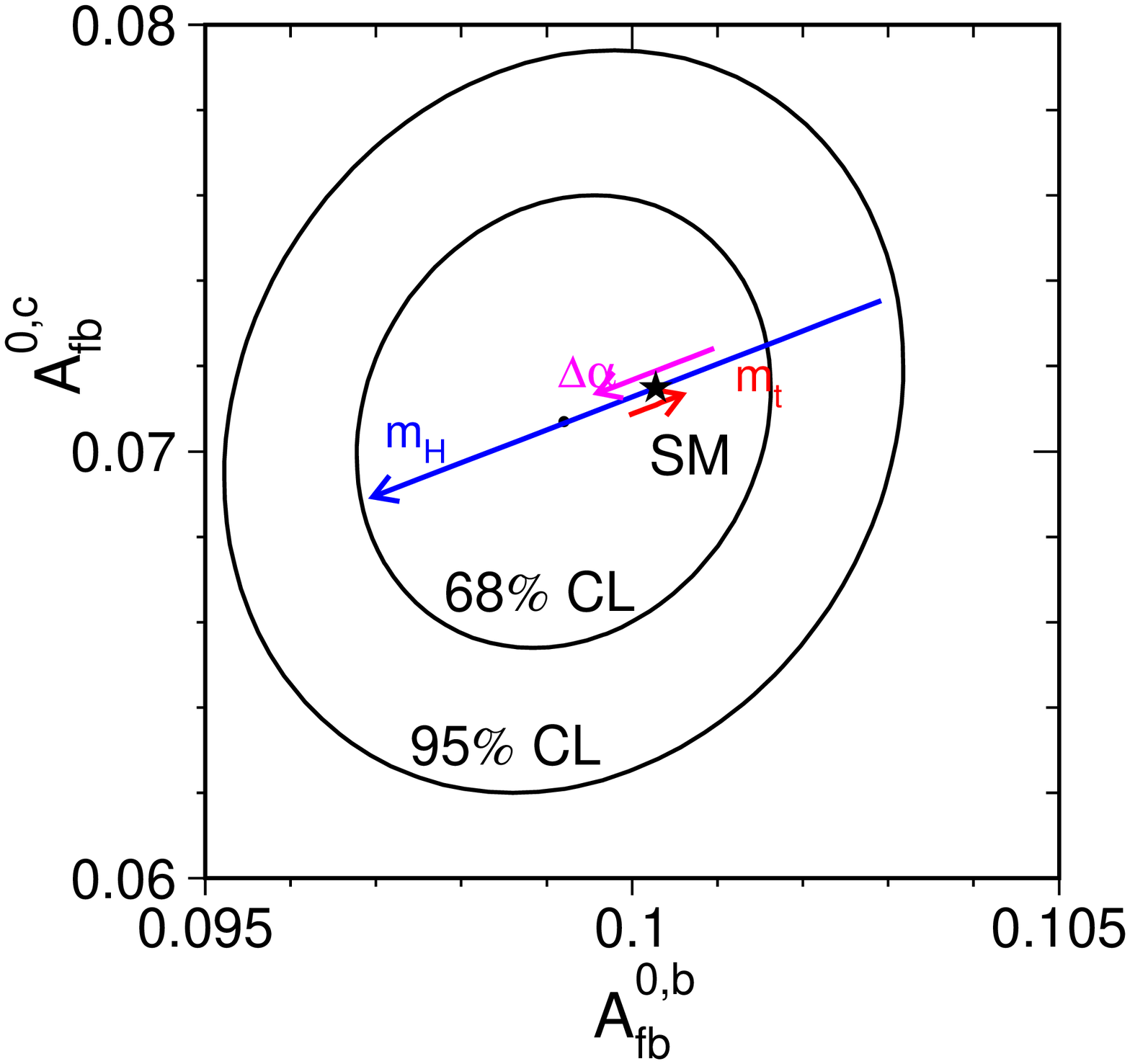}
\hfill
\includegraphics[width=0.49\linewidth]{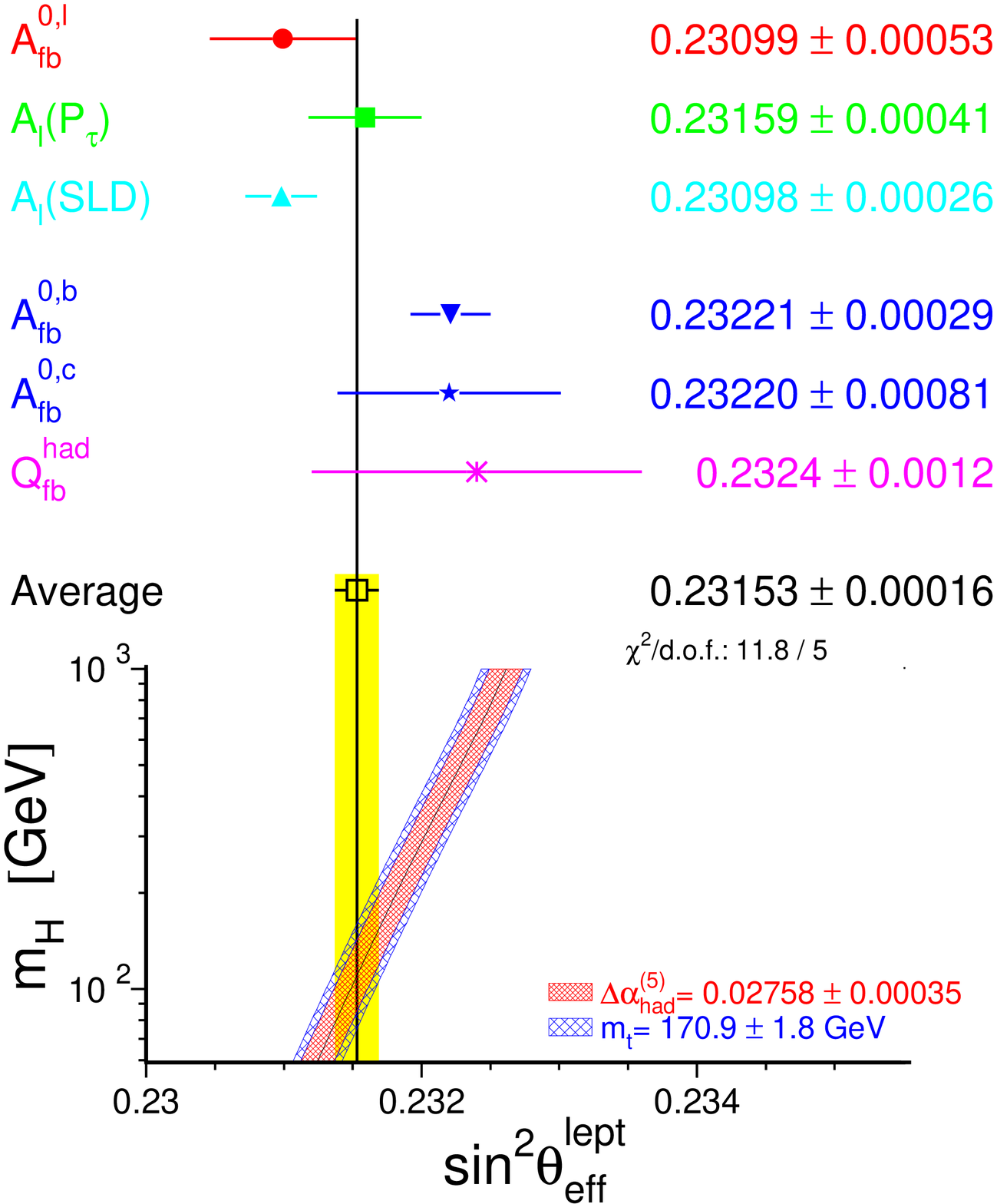}
\vskip -0.5cm
\caption{Left: Contour curves in the $(\Afbzb,\Afbzc)$ plane.  The SM
 expectations are shown as the arrows for $\MT=170.9\pm1.8~\GeV$,
 $\MH=300^{+700}_{-186}~\GeV$ and $\dalhad=0.02758\pm0.00035$.  Right:
 The effective electroweak mixing angle from asymmetry measurements. }
\label{fig:coup:aq}
\end{center}
\end{figure}

The heavy-flavour results at the Z-pole were the last precision
electroweak Z-pole results to become final; details on the various
measurements are given in~\cite{Z-POLE}.  The combination of these
measurements has a rather low $\chi^2$ of 53 for $(105-14)$ degrees of
freedom: all forward-backward asymmetries are very consistent, and
their combination is still statistics limited.  The combined values
for $\Afbzb$ and $\Afbzc$ are compared to the SM expectation in
Figure~\ref{fig:coup:aq} (left), showing that they agree well with the
SM expectation for a medium Higgs-boson mass of a few hundred $\GeV$.

Assuming the SM structure of the effective coupling constants, the
measurements of the various asymmetries are compared in terms of
$\swsqeffl$ in Figure~\ref{fig:coup:aq} (right).  The average of all
$\swsqeffl$ determinations is $\swsqeffl = 0.23153\pm0.00016$, with a
$\chi^2/dof$ of 11.8/5, corresponding to a probability of 3.7\%. The
enlarged $\chi^2/dof$ is solely driven by the two most precise
determinations of $\swsqeffl$, namely those derived from the
measurements of $\Al$ by SLD, dominated by the left-right asymmetry
result, and of $\Afbzb$ at LEP, preferring a low and high Higgs-boson
mass, respectively.  The two measurements differ by 3.2 standard
deviations.

In 1995 the Tevatron experiments CDF and {D\O} discovered the top
quark in proton-antiproton collisions at $1.8~\TeV$ centre-of-mass
energy, by observing the reaction $\Ppp\to\PTT~X,~\PTT\to\Pbb\PWW$.
The results on the mass of the top quark presented at this
conference~\cite{Mtop-EPS}, based on data collected during Run-I
(1992-1996) and the ongoing Run-II (since 2001) are combined by the
Tevatron Electroweak Working~\cite{TEVEWWG07-EPS}: $\MT =
170.9\pm1.1~(stat.)\pm1.5~(syst.)~\GeV$, corresponding to an overall
precision of 1.1\%.

Final results on the mass and width of the W boson from CDF and {D\O}
are available for the complete Run-I data set. First results based on
the Run-II data set are available for $\GW$ from D\O\ and, recently,
for $\MW$ from CDF, with combined results of $\MW=80.429\pm0.039~\GeV$
and $\GW=2.078\pm0.087~\GeV$~\cite{TEV-MW-GW}.  The results on $\MW$
and $\GW$ from the LEP-2 experiments ALEPH, DELPHI, L3 and OPAL are
all final. However, the LEP combined estimation of colour-reconnection
effects based on dedicated studies, used in limiting that uncertainty
in the LEP $\MW$ combination, is still preliminary, and thus is the
LEP combination of $\MW$ and $\GW$: $\MW=80.376\pm0.033~\GeV$ and
$\GW=2.196\pm0.083~\GeV$~\cite{LEP-MW-GW}. The LEP and Tevatron
results are in good agreement; the new preliminary world-average
values are: $\MW=80.398\pm0.025~\GeV$ and $\GW=2.140\pm0.060~\GeV$.
Within the SM, these $\MW$ results points to a low Higgs-boson mass,
as shown in Figure~\ref{fig:sef2-mt-mw} (left), in contrast to
$\Afbzb$.

\section{Combined Electroweak Analysis}
\label{sec:MSM}

Within the framework of the SM, each pseudo observable is calculated
as a function of five relevant input parameters: the running
electromagnetic and strong coupling constants evaluated at the Z pole,
$\alpha_{em}$ and $\aqcd$, and the masses of Z boson, top quark and
Higgs boson, $\MZ$, $\MT$, $\MH$.  Using the Fermi constant $\GF$
allows to calculate the mass of the W boson. The running
electromagnetic coupling is represented by the hadronic vacuum
polarisation $\dalhad$, as it is this contribution which has the
largest uncertainty, $\dalhad=0.02758\pm0.00035$~\cite{BP05}.  The
dependence on $\MT$ and $\MH$ enters through electroweak loop
corrections.  The predictions are calculated with the computer
programs TOPAZ0 and ZFITTER, which incorporate state-of-the-art
calculations of radiative corrections~\cite{TZ}.

\begin{figure}[htbp]
\begin{center}
$ $\vskip -1cm
\includegraphics[width=0.49\linewidth]{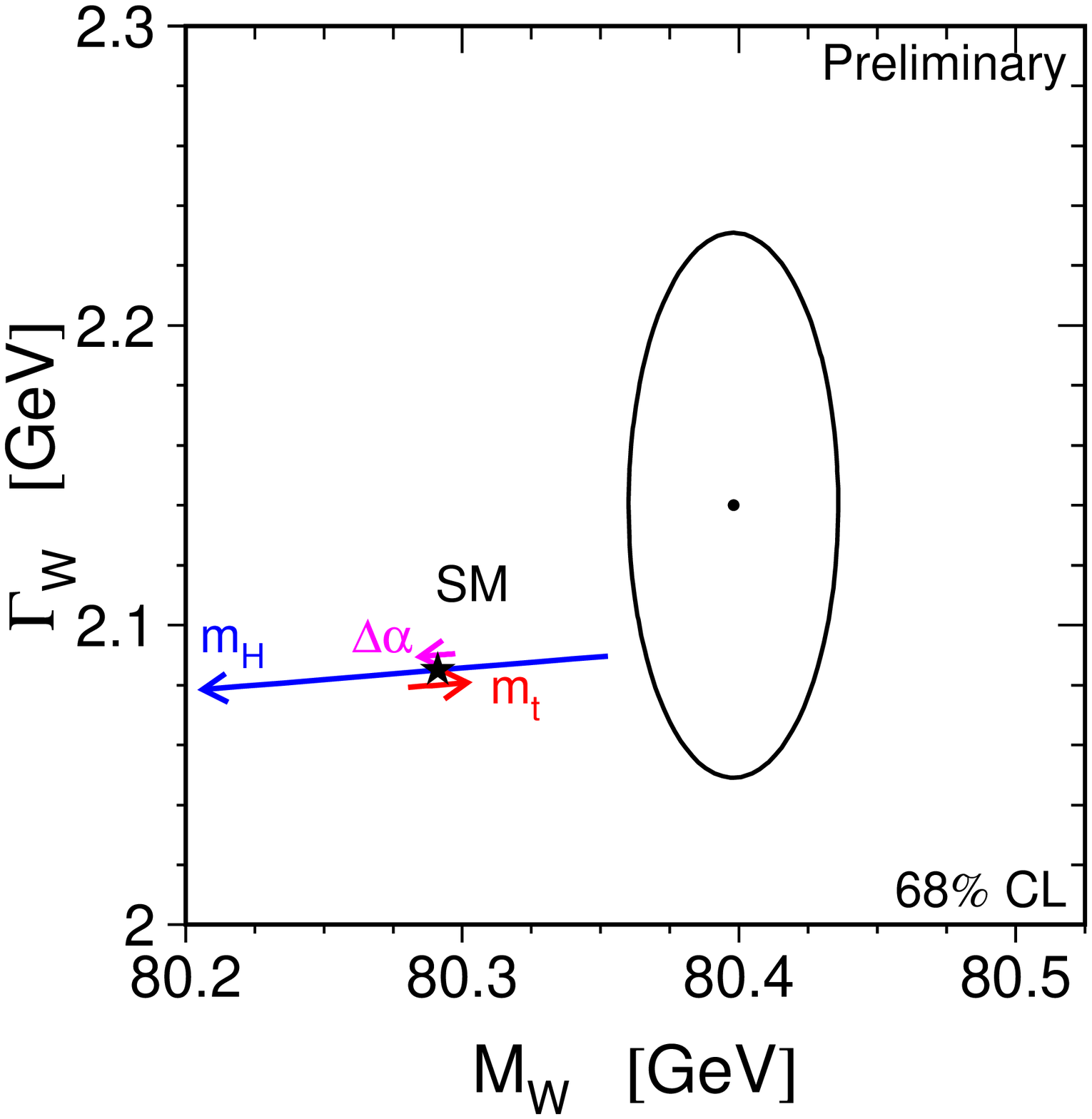}
\hfill
\includegraphics[width=0.49\linewidth]{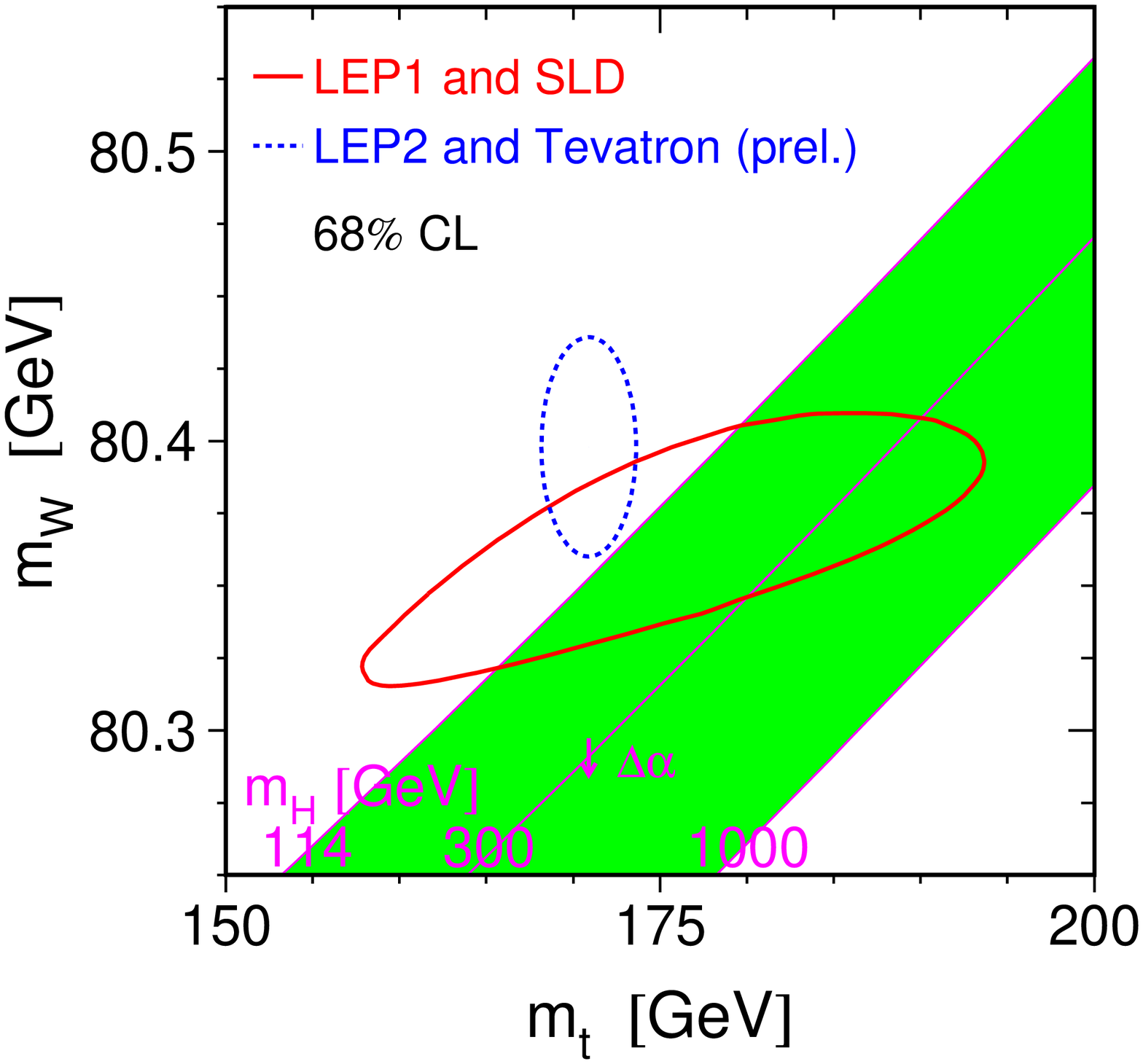}
\vskip -0.9cm
\caption{Left: Contour curves of 68\% C.L. in the $(\MW,\GW)$ plane.
Right: Contour curves of 68\% C.L. in the $(\MT,\MW)$ plane for the
direct measurements and the indirect determinations.  The band shows
the correlation between $\MW$ and $\MT$ expected in the SM. }
\label{fig:sef2-mt-mw}
\end{center}
\end{figure}

Using the Z-pole measurements of SLD and LEP-I, electroweak radiative
corrections are evaluated allowing to predict the masses of top quark
and W boson. The resulting 68\% C.L. contour curve in the $(\MT,\MW)$
plane is shown in Figure~\ref{fig:sef2-mt-mw} (right).  Also shown is
the contour curve corresponding to the direct measurements of both
quantities at the Tevatron and at LEP-II. The two contours overlap,
successfully testing the SM at the level of electroweak radiative
corrections. The diagonal band in Figure~\ref{fig:sef2-mt-mw} (right)
shows the constraint between the two masses within the SM, which
depends on the unknown mass of the Higgs boson, and to a small extent
also on the hadronic vacuum polarisation (small arrow labeled
$\Delta\alpha$).  Both the direct and the indirect contour curves
prefer a low value for the mass of the SM Higgs boson.

\begin{figure}[htbp]
\begin{center}
$ $\vskip -1cm
\includegraphics[width=0.49\linewidth]{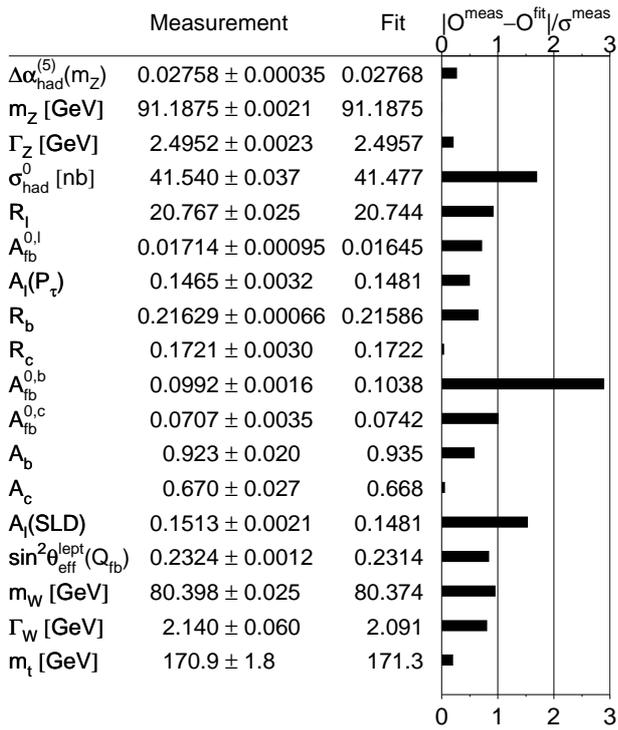}
\hfill
\includegraphics[width=0.49\linewidth]{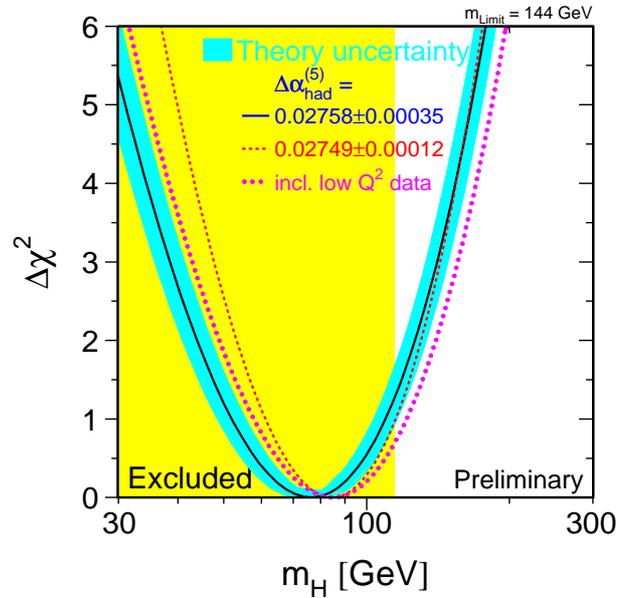}
\vskip -0.9cm
\caption{Left: Pulls of the measurements used in the global SM
  analysis. Right: $\Delta\chi^2$ curve as a function of $\MH$.  Also
  shown are the curves using a theory-driven evaluation of $\dalhad$,
  or including the low-$Q^2$ measurements in the analysis.  }
\vskip -1cm
$ $
\label{fig:pulls-blue}
\end{center}
\end{figure}

The best constraint on $\MH$ is obtained by analysing all data.  This
global fit has a $\chi^2$ of 18.2 for 13 degrees of freedom,
corresponding to a probability of 15.1\%. The pulls of the 18
measurements fitted are shown in Figure~\ref{fig:pulls-blue} (left).
The single largest contribution to the $\chi^2$ arises from the
$\Afbzb$ measurement discussed above, with a pull of 2.9.  The fit
yields $\MH = 76^{+33}_{-24}~\GeV$, a 37\% constraint on $\MH$, which
corresponds to a one-sided 95\% C.L. upper limit on $\MH$ of
$144~\GeV$ including the theory uncertainty, as shown in
Figure~\ref{fig:pulls-blue} (right). This limit increases to
$182~\GeV$ when including the LEP-2 direct-search limit of
$114.4~\GeV$~\cite{LEP-HIGGS} in the analysis.

The fitted $\MH$ is strongly correlated with the fitted hadronic
vacuum polarisation (correlation of $-0.54$) and the fitted top-quark
mass ($+0.39$).  The strong correlation with $\MT$ implies a shift of
15\% in $\MH$ if the measured $\MT$ changes by $2~\GeV$.  Thus a
precise measurement of $\MT$ is very important.  Also shown are the
$\chi^2$ curves obtained with the more precise but theory-driven
evaluation of $\dalhad$~\cite{YNDURAIN}, yielding a correlation of
only $-0.2$ with $\MH$, or including the results obtained in low-$Q^2$
interactions: atomic parity violation~\cite{APV-Caesium}, Moller
scattering~\cite{E-158}, and NuTeV's measurement of deep-inelastic
lepton-nucleon scattering~\cite{NuTeV}; with the two former
measurements in agreement with the expectations but the latter
differing by 3 standard deviations.  Both analyses yield nearly the
same upper limits on $\MH$.

The theoretical uncertainty on the SM calculations of the observables
is shown as the thickness of the blue band. It is dominated by the
uncertainty in the calculation of the effective electroweak mixing
angle, where a completed two-loop calculation is needed.  The shaded
part in Figure~\ref{fig:pulls-blue} (right) shows the $\MH$ range up
to $114.4~\GeV$ excluded by the direct search for the Higgs boson at
95\% confidence level~\cite{LEP-HIGGS}. Even though the minimum of the
$\chi^2$ curve lies in the excluded region, the uncertainties on the
fitted Higgs mass are such as that the results are well compatible.

\section{Conclusions and Outlook}

Over the last 2 decades many experiments have performed a wealth of
measurements with unprecedented precision in high-energy particle
physics. These measurements test all aspects of the SM of particle
physics, and many of them show large sensitivity to electroweak
radiative corrections, and point to a light SM Higgs boson.  Most
measurements agree well with the expectations as calculated within the
framework of the SM, successfully testing the SM at Born and loop
level. There are two ``3 standard deviations effects'', namely the
spread in the various determinations of the effective electroweak
mixing angle, within the SM analysis disfavouring the measurement of
$\Afbzb$, and NuTeV's result, most pronounced when interpreted in
terms of the on-shell electroweak mixing angle.  For the future,
precise theoretical calculations including theoretical uncertainties
are needed, in particular a completed two-loop calculation for the
effective electroweak mixing angle and a NLO reanalysis of the NuTeV
measurement.  Experimentally, the next few years will bring further
improvements in the measurements of W-boson and top-quark masses,
allowing to constrain $\MH$ to 28\% (Tevatron/LHC) and even 16\%
(ILC/GigaZ).  Of course, the discovery of the Higgs boson is eagerly
awaited, with a measurement of its mass to sub-$\GeV$ precision and of
other properties.

\vfill
\noindent
{\bf Acknowledgements}

It is a delight to thank my colleagues of the LEP and Tevatron
electroweak working groups, members of the ALEPH, CDF, DELPHI, D\O,
OPAL and SLD experiments, as well as Tord Riemann and Georg Weiglein
for valuable discussions.


\vfill


\end{document}